\documentclass[5p,twocolumn,times,number]{elsarticle}

\usepackage{lineno}
\usepackage{graphicx}
\usepackage{caption}
\usepackage{subcaption}
\usepackage{siunitx}

\usepackage[belowskip=-15pt,aboveskip=0pt]{caption}

\usepackage[compact]{titlesec}
\titlespacing{\section}{0pt}{*0}{*0}

\journal{Elsevier}

\begin{document}

\begin{frontmatter}


\title{Optimized MPGD-based Photon Detectors for high momentum particle identification at the Electron-Ion Collider}

\author[add1]{J.~Agarwala\corref{cor}}
\ead{jinky.agarwala@ts.infn.it}
\author[add2]{F.~Bradamante}
\author[add2]{A.~Bressan}
\author[add2]{C.~Chatterjee}
\author[add2]{P.~Ciliberti}
\author[add1]{S.~Dalla Torre}
\author[add1]{S.~Dasgupta}
\author[add1]{M.~Gregori}
\author[add1]{S.~Levorato}
\author[add2]{A.~Martin}
\author[add1]{G.~Menon}
\author[add1]{F.~Tessarotto}
\author[add1]{Y.~Zhao}

\cortext[cor]{Corresponding author}

\address[add1]{INFN Trieste, Trieste, Italy}
\address[add2]{University of Trieste and INFN Trieste, Trieste, Italy}
\begin{abstract} 
Particle IDentification (PID) is a central requirement of the experiments at the future EIC. Hadron PID at high momenta by RICH techniques requires the use of low density gaseous radiators, where the challenge is the limited length of the radiator region available at a collider experiment. By selecting a photon wavelength range in the far UV domain, around 120 nm, the number of detectable photons can be increased. Ideal sensors are gaseous Photon Detectors (PD) with CsI photocathode, where the status of the art is represented by the MPGD-based PDs at COMPASS RICH. Detector optimization is required for the application at EIC.

Here we report about a dedicated prototype where the sensor pad-size has been reduced to preserve the angular resolution. A new DAQ system based on the SRS readout electronics has been developed for the laboratory and test beam studies of the prototype.
\end{abstract}


\end{frontmatter}


\section{Introduction}
\label{S:1}

Quantum Chromodynamics, the theory of strong interaction and hadronic matter, acts as a bridge between fundamental particles and nuclei. The Electron-Ion Collider (EIC) \cite{Accardi:2016}, is an ideal accelerator facility with unprecedented luminosity, wide energy range and beams of spin-polarized electrons colliding off beams of polarized nucleons or unpolarized nuclei from deuterium to uranium to explore this bridge. It has been recommended by the U.S. Nuclear Science Advisory Committee 2015 Long Range Plan. PID is one of the most demanding sectors for the experiments at EIC: in particular, hadron PID relies on the application of Ring Imaging CHerenkov (RICH) techniques. The challenge of limited length of the radiator compatible with the collider experiments imposes specific criteria for the RICH:
\begin{itemize}
\item High momenta hadron PID requires low-density radiator gas. A concrete option to increase the number of detectable photons is to choose the photon wavelength range in the far UV domain, around 120 nm. In this domain, gaseous PDs with CsI photocathode are ideal sensors.
\item A shorter radiator implies a RICH design with shorter focal length. The strategy to preserve the resolution in the measurement of Cherenkov angle is by an improved space resolution of the photon detectors.
\end{itemize}

\section{The state of the art}
\label{S:2}
MPGD-based Photon Detectors (PD), sensible in the UV range are successfully in operation for COMPASS RICH-1 at CERN. They cover a total 1.4 m$^2$ area and represent the status of the art. Four chambers of PDs have been implemented during 2016 RICH-1 upgrade \cite{Agarwala:2018}. Each chamber (60 $\times$ 60 cm$^2$) consists of two THGEM layers coupled to a resistive MicroMegas (MM) with a readout anode segmented in 8 $\times$ 8 mm$^2$ pads.
\section{The new prototype PD}
\label{S:3}
\begin{figure}[!h]
\centering\includegraphics[width=0.9\linewidth]{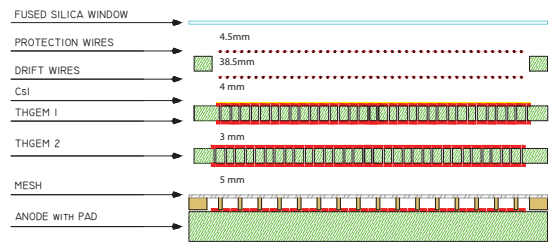}
\caption{Sketch of the hybrid single photon detector: two staggered THGEM
layers are coupled to a resistive bulk MM. Image not to scale.}
\label{fig:f1}
\end{figure}

\begin{figure}[!htbp]
  \begin{subfigure}[b]{0.2\textwidth}
    \includegraphics[width=\textwidth]{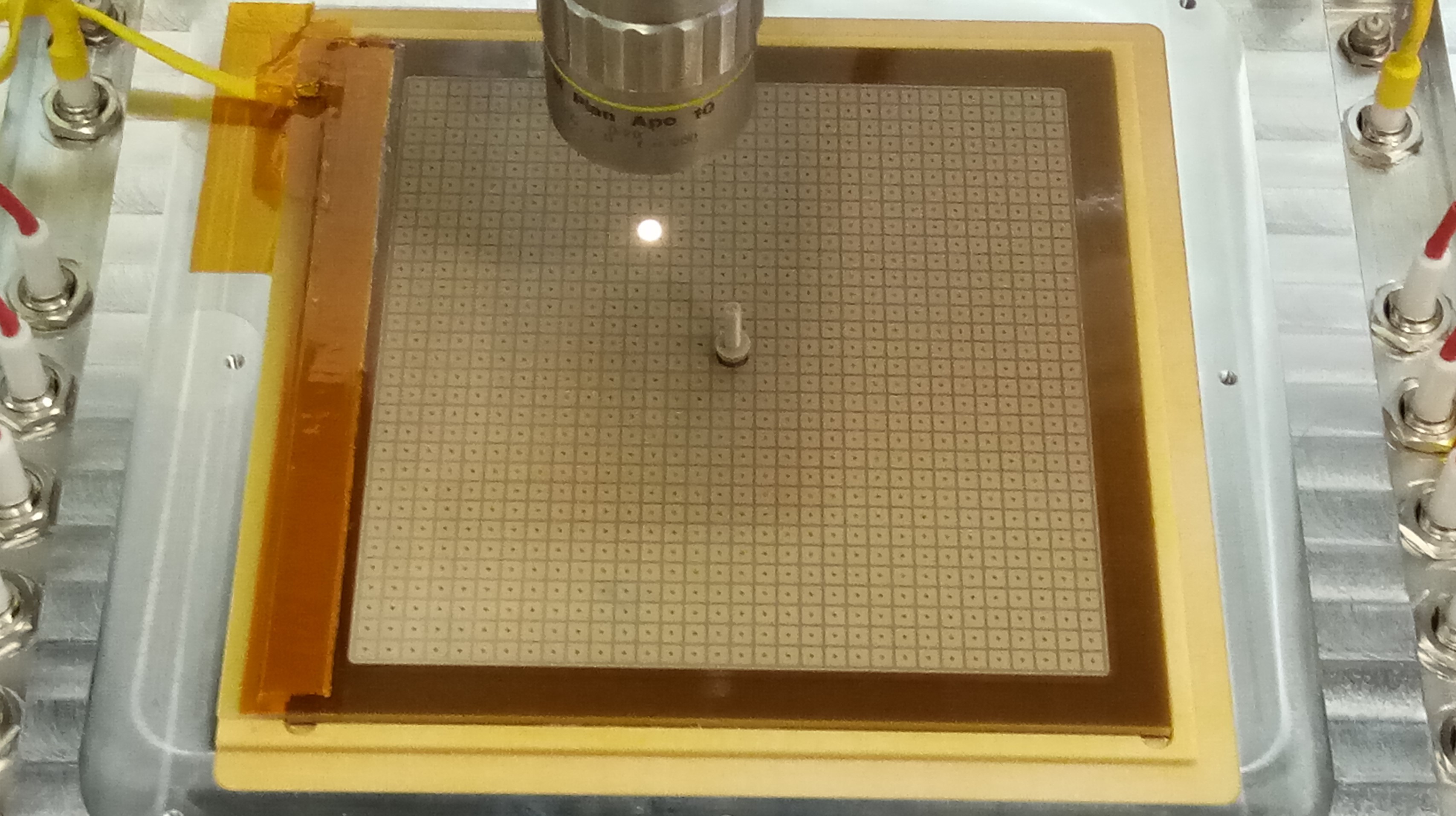}
    \label{fig:f2a}
  \end{subfigure}
  \hfill
  \begin{subfigure}[b]{0.1\textwidth}
    \includegraphics[width=\textwidth]{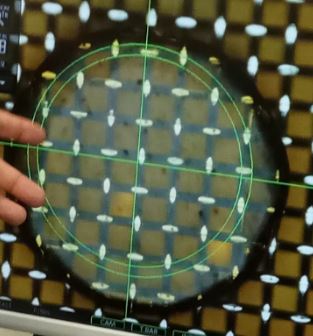}
    \label{fig:f2b}
  \end{subfigure}
  \hfill
  \begin{subfigure}[b]{0.16\textwidth}
    \includegraphics[width=\textwidth]{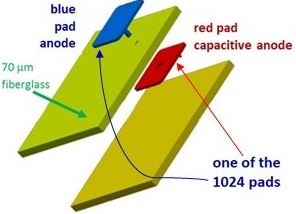}
    \label{fig:f2c}
  \end{subfigure}
  \caption{Left) Picture of the MM stage of the prototype chamber. Centre) Picture of one of the pillars supporting the micromesh. Right) Sketch of the capacitive coupled readout pad.}
  \label{fig:f2}
\end{figure}

An evolution of COMPASS PDs has been designed and a prototype chamber has been built at INFN Trieste. Two THGEMs are followed by one bulk MM (Fig.\ref{fig:f1}). The active area of the chamber is 10 $\times$ 10 cm$^2$ (Fig.\ref{fig:f2}-Left). MM have 18 $\mu$m woven stainless steel wire mesh with 63 $\mu$m pitch. A square array of pillars (Fig.\ref{fig:f2}-Center) provides the correct geometry of the MM. The readout anode is highly segmented in 1024 square pads each having 3 $\times$ 3 mm$^2$ area. The anode pads, facing the micromesh, are individually powered and individually protected by a 0.5 G\si{\ohm} resistor. Eight resistor cards distribute the HV to 128 pads each. The signals induced via capacitive couplings (Fig.\ref{fig:f2}-Right) in the buried pads are readout by the electronics. The modular design of the readout does not exceed the active area to guarantee the extendibility of the design to larger detector size.
\section{Readout system and DAQ}
\label{S:4}
\subsection{Hardware}
Signal collecting and processing are based on the RD51 Scalable Readout System (SRS) \cite{Martoiu:2013}. The Front End is based on the APV25 analog chip. Each of the eight APV chips reads 128 pads at a frequency of 40 MHz. The analog data is sent to the ADC card managed by an FPGA. The digitized data is then sent, using the UDP protocol via Ethernet port, to a PC for data collection and storage.
\begin{figure}[!htbp]
  \begin{subfigure}[b]{0.41\textwidth}
    \includegraphics[width=\textwidth]{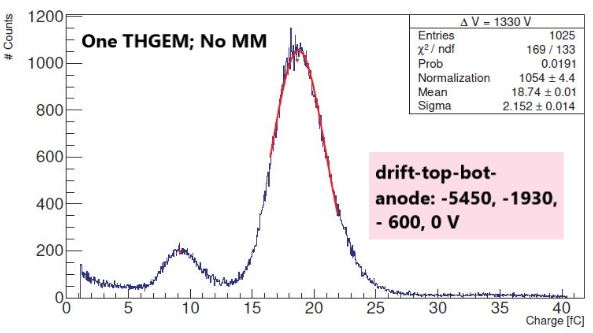}
    \label{fig:f3a}
  \end{subfigure}
  \hfill
  \begin{subfigure}[b]{0.41\textwidth}
    \includegraphics[width=\textwidth]{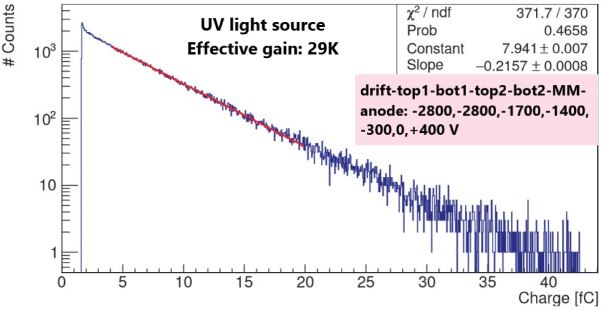}
    \label{fig:f3b}
  \end{subfigure}
  \caption{Spectra obtained with a THGEM alone; top) X-ray from a $^{55}$Fe source; bottom) single photons.}
  \label{fig:f3}
\end{figure}

\subsection{Software (Raven DAQ and Raven CODE)}
An original LabView based novel Data AcQuisition (DAQ) system, called Raven, has been developed to control the data acquisition and increase the acquisition trigger rate. The settings and configuration of APV25 chips are performed via Raven DAQ, which also provides fast data processing taking care of pedestal subtraction and zero suppression to compact data. The data are handled offline by the RavenCode, written in C++ (enriched with a ROOT interface), in 3 steps: RavenPedestal performing pedestal subtraction, RavenDecoder reconstructing the geometrical address of each hit channel and RavenAnalyzer performing data analysis. Raven has extremely user friendly GUI.

\section{Experimental data and observations}
\label{S:5}
\begin{figure}[!h]
\centering\includegraphics[width=0.65\linewidth]{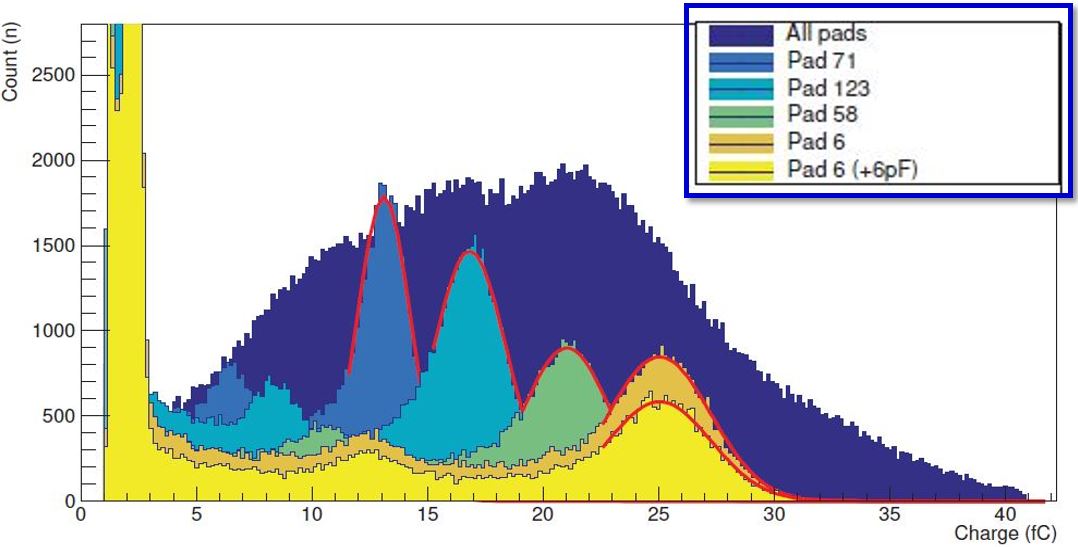}
\caption{$^{55}$Fe amplitude spectra collected by different pads using the MM multiplication stage only.}
\label{fig:f4}
\end{figure}

\begin{figure}[!h]
\centering\includegraphics[width=0.8\linewidth]{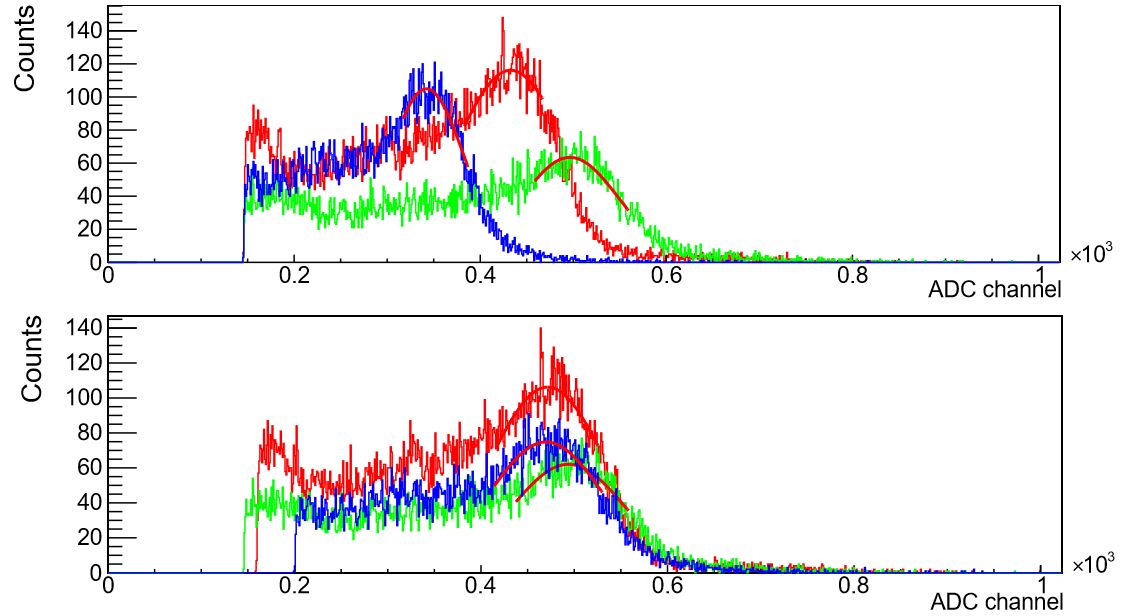}
\caption{Excited Cu amplitude spectra collected with three different pads; top) raw spectra. bottom) corrected spectra accounting for the different capacitance of the pads.}
\label{fig:f5}
\end{figure}

The various multipliers of the detector exhibit good electrical stability and gain performance. Proper spectra are collected with the THGEMs  (thickness 0.4 mm, pitch 0.8 mm, hole diameter 0.4 mm, no rim) both illuminated with  a $^{55}$Fe source (Fig.\ref{fig:f3}-Top) and by single photons from an UV LED (Fig.\ref{fig:f3}-Bottom). A problematic aspect appeared using the MM prototype: different pads provide different signal amplitude distributions with variations up to a factor 2 even between adjacent pads (Fig.\ref{fig:f4}). The source of the non-uniformity could be identified in the parasitic capacitance, different pad by pad, in the first version of the anode PCB. Correcting the amplitude spectra by the measured effect of the parasitic capacitance (Fig.\ref{fig:f5}) uniformity can be restored. A new version of the anode PCB has been designed. It will be realized for a future second version of the prototype.
\section{Conclusions}
\label{S:6}
A new MPGD-based PD with highly segmented anode is being developed to match the challenging PID needs at EIC. A first prototype has been tested in laboratory; it exhibits good performance, apart from the gain non-uniformity of the MM stage. The source of the non-uniformity is not related to the detector architecture: it will be overcome in a next version of the prototype. In the context of the R\&D work, an original DAQ system has been developed for the SRS readout allowing for fast and efficient data collection and reconstruction.

\section*{Acknowledgments}
This work is partially supported by the H2020 project AIDA-2020, GA no. 654168. J. Agarwala is supported by an ICTP TRIL fellowship.


\end{document}